\documentclass[12pt]{article}
\usepackage{amsfonts}
\usepackage{amsmath}
\usepackage{amssymb}
\usepackage{graphicx}
\usepackage{color}
\usepackage[all, knot]{xy}
\usepackage{tikz}

\usepackage{ulem}

\usepackage[utf8]{inputenc}
\usepackage{epstopdf}
\usepackage[footnotesize]{caption}
\usepackage{amsthm}
\usepackage{enumitem}
\usepackage{mathrsfs}

\usepackage[margin=3cm]{geometry}

\def \be {\begin{equation}}
\def \ee {\end{equation}}
\def \bea {\begin{eqnarray}}
\def \eea {\end{eqnarray}}
\def \nn {\nonumber}

\def \rr {\raise.35ex\hbox{\small $\prime$}\kern-.17em{\mbox{\large $\imath$}}}

\def \dels {\partial\kern-.6em /\kern.1em}
\def \As {{A\kern-.5em / \kern.5em}}
\def \Ds {D\kern-.7em / \kern.5em}

\def \ks {k\kern-.5em /}
\def \ls {l\kern-.5em /}

\newcommand{\ci}[1]{}

\newcommand{\ba}{\begin{eqnarray}}
\newcommand{\ea}{\end{eqnarray}}
\newcommand{\bal}{\begin{align}}
\newcommand{\eal}{\end{align}}
\newcommand{\bay}[1]{\left(\begin{array}{#1}}
\newcommand{\eay}{\end{array}\right)}

\newcommand{\hide}[1]{}

\DeclareMathOperator{\sech}{sech}

\newlist{axioms}{enumerate}{2}
\setlist[axioms,1]{label=\textbf{A\arabic{axiomsi}.}, ref=A\arabic{axiomsi}}
\setlist[axioms,2]{label=\textbf{A\arabic{axiomsi}\rlap{\myEnumCounter{axiomsii}}.},%
                   ref=A\arabic{axiomsi}\myEnumCounter{axiomsii},%
                   align=parleft,%
                   leftmargin=0em,%
                   itemsep=1.4ex,%
                   before={\stepcounter{axiomsi}}}

  \usetikzlibrary{decorations.markings}

\begin{document}
\begin{titlepage}
\begin{flushright}
NORDITA 2019-102 
\end{flushright}

\begin{center}

\textbf{\LARGE
Quantum Correction of the\\
 Wilson Line and Entanglement Entropy\\ in the\\
  Pure AdS$_3$ Einstein Gravity Theory
\vskip.3cm
}
\vskip .5in
{\large
Xing Huang$^{a, b}$ \footnote{e-mail address: xingavatar@gmail.com},
Chen-Te Ma$^{c, d, e}$ \footnote{e-mail address: yefgst@gmail.com}, 
and Hongfei Shu$^{f, g}$ \footnote{e-mail address: hongfei.shu@su.se}
\\
\vskip 1mm
}
{\sl
$^a$
Institute of Modern Physics, Northwest University, Xi'an 710069, China.
\\
$^b$
Shaanxi Key Laboratory for Theoretical Physics Frontiers, Xi'an 710069, China.
\\
$^c$
Guangdong Provincial Key Laboratory of Nuclear Science,\\
 Institute of Quantum Matter,
South China Normal University, Guangzhou 510006, Guangdong, China.
\\
$^d$
School of Physics and Telecommunication Engineering,\\
 South China Normal University, Guangzhou 510006, Guangdong, China.
\\
$^e$
The Laboratory for Quantum Gravity and Strings,\\
 Department of Mathematics and Applied Mathematics,\\
University of Cape Town, Private Bag, Rondebosch 7700, South Africa.
\\
$^f$
Nordita, KTH Royal Institute of Technology and Stockholm University,\\
 Roslagstullsbacken 23, SE-106 91 Stockholm, Sweden.
\\
$^g$
Department of Physics, Tokyo Institute of Technology, Tokyo, 152-8551, Japan.
}
\\
\vskip 1mm
\vspace{40pt}
\end{center}
\newpage
\begin{abstract}
We calculate the expectation value of the Wilson line in the pure AdS$_3$ Einstein gravity theory and also the entanglement entropy in the boundary theory. Our one-loop calculation of entanglement entropy shows a shift of the central charge 26.
Finally, we show that the Wilson line provides the equivalent description to the boundary entanglement entropy. This equivalence leads to a concrete example of the building of ``minimum surface=entanglement entropy''.
\end{abstract}
\end{titlepage}

\section{Introduction}
\label{sec:1}
\noindent
The holographic principle states that physical degrees of freedom in quantum gravity theory is fully encoded by the boundary \cite{tHooft:1993dmi}. Because Einstein gravity theory \cite{Kraus:2006wn} loses renormalizability, a direct study in quantum gravity theory is not easy. The holographic principle provides the boundary perspective to study quantum gravity theory. In this direction, the most well-studied is the Anti-de Sitter/ Conformal Field Theory (AdS/CFT) correspondence \cite{Oblak:2016eij}.
\\

\noindent
The AdS/CFT correspondence was motivated and conjectured by the ultraviolet complete theory, string theory. Therefore, this conjecture is quite realizable. The AdS black hole solution \cite{Banados:1998gg} also provides an application to condensed matter systems \cite{Ma:2018efs}.
A new application of the AdS/CFT correspondence is the equivalence between entanglement entropy of CFT \cite{Calabrese:2009qy} and the codimension-two minimum surface at a given time slice in the AdS background \cite{Ryu:2006bv}. The replica trick \cite{Holzhey:1994we} is the most general method for computing entanglement entropy, but it is still hard to obtain an exact solution. The holographic method \cite{Takayanagi:2012kg} provides usefulness to studying entanglement entropy in strongly coupled CFT \cite{Nishioka:2009un}. The proposal becomes more realizable by using the replica trick in the bulk gravity theory to reach the same conclusion \cite{Lewkowycz:2013nqa}. Because the entanglement entropy needs to be defined by the decomposition, the gravity and gauge theories should suffer from the non-gauge invariant cutting. Nevertheless, borrowing the von Neumann algebra gives an interpretation for doing a partial trace operation without breaking the gauge symmetry in each sub-region \cite{Ma:2015xes}. Hence this holographic study gives a concretely useful application to the AdS/CFT correspondence.
\\

\noindent
What people are mostly interested in the holographic principle is pure Einstein gravity theory. Nevertheless, people cannot probe a quantum region due to the well-known issue, renormalizability. Nowadays, the closest route is the AdS$_3$ Einstein gravity theory \cite{Carlip:2005zn}, defined by a gauge theory \cite{Witten:1988hc}, not by a metric formulation. The gauge formulation is the SL(2) Chern-Simons gravity theory \cite{Witten:1988hc}. This theory can be quantized and is renormalizable \cite{Elitzur:1989nr}. The gauge formulation is equivalent to the metric formulation only up to the classical and perturbation level \cite{Witten:2007kt}. Two formulations are not equivalent exactly. This gauge formulation \cite{Vasiliev:2001ur} is also interesting for the simple extension of higher spins \cite{Bekaert:2005vh} with a unified study from the SL($M$) groups \cite{Ammon:2012wc}.
\\

\noindent
Since the three-dimensional Einstein gravity theory does not have the local gravitation fluctuation, a direct derivation of the boundary theory is possible. The original derivation points to CFT$_2$, the Liouville theory \cite{Coussaert:1995zp}. This theory does not have a normalizable vacuum. It contradicts a known truth. This contradiction goes away after one found two-dimensional Schwarzian theory \cite{Cotler:2018zff}. In a quantum regime, this boundary theory breaks conformal symmetry in the sense that modular invariance is absent.
\\

\noindent
The Wilson line \cite{Ammon:2013hba} provides the universal contribution \cite{Brown:1986nw} to the holographic entanglement entropy at the classical level \cite{deBoer:2013vca}. The central question that we would like to address in this letter is the following: {\it What is the quantum deformation of the minimum surface?} Since the minimum surface is only defined at the classical level, we would like to study the quantum deformation using the computable gauge formulation. 
\\

\noindent
We precisely compute \cite{Szabo:1996md} entanglement entropy for one-interval using the $n$-sheet partition function \cite{Huang:2016bkp}. Then we show that the bulk operator, Wilson line, is exactly dual to the boundary entanglement entropy. We provide the quantum deformation of the minimum surface \cite{Ma:2016deg} from the Wilson line with the dual of entanglement entropy \cite{Ma:2016pah}.

\section{SL(2) Chern-Simons Gravity Theory}
\label{sec:2}
\noindent
The action of the SL(2) Chern-Simons gravity theory is given by \cite{Witten:1988hc}
\bea
&&S_{\mathrm{G}}
\nn\\
&=&\frac{k}{2\pi}\int d^3x\ \epsilon^{tr\theta}\mathrm{Tr}\bigg(A_tF_{r\theta}-\frac{1}{2}\big(A_r\partial_tA_{\theta}-A_{\theta}\partial_tA_r\big)\bigg)
\nn\\
&&-\frac{k}{2\pi}\int d^3x\ \epsilon^{tr\theta}\mathrm{Tr}\bigg(\bar{A}_t\bar{F}_{r\theta}-\frac{1}{2}\big(\bar{A}_r\partial_t\bar{A}_{\theta}-\bar{A}_{\theta}\partial_t\bar{A}_r\big)\bigg)
\nn\\
&&-\frac{k}{4\pi}\int dtd\theta\ \mathrm{Tr}(A_{\theta}^2)
\nn\\
&&-\frac{k}{4\pi}\int dtd\theta\ \mathrm{Tr}(\bar{A}_{\theta}^2),
\eea
in which we assume that the boundary conditions of the gauge fields $A$ and $\bar{A}$ are: $A_-\equiv A_t-A_{\theta}=0$ and $\bar{A}_+=A_t+A_{\theta}=0$.
The variable $k$ is defined by $l/(4G_3)$, where $1/l^2\equiv-\Lambda$. The cosmological constant is denoted by $\Lambda$, and the three-dimensional gravitational constant is denoted by $G_3$. 
The gauge fields are defined by the vielbein $e_{\mu}$ and spin connection $\omega_{\mu}$:
\bea
A_{\mu}&\equiv& A_{\mu}^aJ_a\equiv J_a\bigg(\frac{1}{l}e^a_{\mu}+\omega_{\mu}^a\bigg), 
\nn\\
\bar{A}_{\nu}&\equiv& \bar{A}_{\nu}^a\bar{J}_a\equiv\bar{J}_a\bigg(\frac{1}{l} e^a_{\nu}-\omega^a_{\nu}\bigg),
\eea
in which the Lie algebra indices are labeled by $a$, and the indices are raised or lowered by $\eta\equiv\mathrm{diag}(-1,1,1)$. 
The spacetime indices are labeled by $\mu$ and $\nu$. 
This bulk terms in this theory are equivalent to the Chern-Simons theory up to a boundary term. The measure in this gravitation theory is $\int {\cal D}A{\cal D}\bar{A}$.
\\

\noindent
The SL(2)$\times$SL(2) generators are given by the followings:
\bea
J_0&\equiv&\begin{pmatrix}
0&-\frac{1}{2}\\
\frac{1}{2}&0
\end{pmatrix}, \qquad 
J_1\equiv\begin{pmatrix}
0&\frac{1}{2}\\
\frac{1}{2}&0
\end{pmatrix}, 
\nn\\
J_2&\equiv&\begin{pmatrix}
\frac{1}{2}&0\\
0&-\frac{1}{2}
\end{pmatrix},
\nn\\
\bar{J}_0&\equiv&\begin{pmatrix}
0&-\frac{1}{2}\\
\frac{1}{2}&0
\end{pmatrix}, \qquad 
\bar{J}_1\equiv\begin{pmatrix}
0&-\frac{1}{2}\\
-\frac{1}{2}&0
\end{pmatrix}, 
\nn\\
\bar{J}_2&\equiv&\begin{pmatrix}
\frac{1}{2}&0\\
0&-\frac{1}{2}
\end{pmatrix}.
\eea
The generators satisfy the algebra: 
\bea
\lbrack J^a, J^b\rbrack&=&\epsilon^{abc}J_c, \qquad \mathrm{Tr}\big(J^aJ^b\big)=\eta^{ab}/2; 
\nn\\
\lbrack \bar{J}^a, \bar{J}^b\rbrack&=&-\epsilon^{abc}\bar{J}_c, \qquad \mathrm{Tr}\big(\bar{J}^a \bar{J}^b\big)=\eta^{ab}/2.
\eea
\\

\noindent
The AdS$_3$ geometry is $ds_3^2=-(r^2+1)dt^2+dr^2/(r^2+1)+r^2d\theta^2$,
in which the ranges of coordinates are defined by that $-\infty< t<\infty$, $0<r<\infty$, and $0<\theta\le 2\pi$. We also choose the unit $\Lambda=-1$. 
The metric is defined by the vielbein $g_{\mu\nu}\equiv 2\cdot\mathrm{Tr}(e_{\mu}e_{\nu})$.
\\

\noindent
We substitute the solution ($F_{r\theta}=0$ and $\bar{F}_{r\theta}=0$) into the action, and use the asymptotic boundary condition to get: 
\bea
g^{-1}_{\mathrm{SL(2)}}\partial_{\theta}g_{\mathrm{SL(2)}}|_{r\rightarrow\infty}
=A_{\theta}|_{r\rightarrow\infty}, \qquad 
\bar{g}^{-1}_{\mathrm{SL(2)}}\partial_{\theta}\bar{g}_{\mathrm{SL(2)}}|_{r\rightarrow\infty}
=\bar{A}_{\theta}|_{r\rightarrow\infty},
\eea
 which are fixed by the metric of AdS boundary.
Using the SL(2) transformations: 
\bea
g_{\mathrm{SL(2)}}&=&
\begin{pmatrix}
1& 0
\\
F& 1
\end{pmatrix}
\begin{pmatrix}
\lambda & 0
\\
0& \frac{1}{\lambda}
\end{pmatrix}
\begin{pmatrix}
1 &\Psi
\\
0& 1
\end{pmatrix}, 
\nn\\
\bar{g}_{\mathrm{SL(2)}}&=&
\begin{pmatrix}
1& -\bar{F}
\\
0& 1
\end{pmatrix}
\begin{pmatrix}
\frac{1}{\bar{\lambda}} & 0
\\
0& \bar{\lambda}
\end{pmatrix}
\begin{pmatrix}
1 &0
\\
-\bar{\Psi}& 1
\end{pmatrix},
\eea
we obtain the boundary conditions: $\lambda^2\partial_{\theta}F=2r$, $\partial_{\theta}^2 F/\partial_{\theta} F=-4r\Psi$, $\bar{\lambda}^2\partial_{\theta}\bar{F}=2r$, and $\partial_{\theta}^2 \bar{F}/\partial_{\theta} \bar{F}=-4r\bar{\Psi}$,
which eventually give the boundary theory, two-dimensional Schwarzian theory \cite{Cotler:2018zff},
\bea
&&S_{\mathrm{G}}
\nn\\
&=&\frac{k}{2\pi}\int dtd\theta\ \bigg(\frac{3}{2}\frac{(\partial_-\partial_{\theta}F)(\partial_{\theta}^2F)}{(\partial_{\theta}F)^2}-\frac{\partial_-\partial_{\theta}^2F}{\partial_{\theta}F}
\bigg)
\nn\\
&&-\frac{k}{2\pi}\int dtd\theta\ \bigg(\frac{3}{2}\frac{(\partial_+\partial_{\theta}\bar{F})(\partial_{\theta}^2\bar{F})}{(\partial_{\theta}\bar{F})^2}-\frac{\partial_+\partial_{\theta}^2\bar{F}}{\partial_{\theta}\bar{F}}
\bigg),
\eea
where 
\bea
x^+\equiv t+\theta, \qquad x^-\equiv t-\theta,
\eea
\bea
\partial_+=\frac{1}{2} \partial_t+\frac{1}{2}\partial_{\theta}, \qquad \partial_-=\frac{1}{2}\partial_t-\frac{1}{2}\partial_{\theta}.
\eea
The measure is $\int dFd\bar{F}\ \big(1/(\partial_{\theta}F\partial_{\theta}\bar{F})\big)$.

\section{Entanglement Entropy in\\
 the Two-Dimensional Schwarzian Theory}
\label{sec:3}
\noindent
We first write down the boundary theory on the sphere manifold \cite{Cotler:2018zff} and then calculate the $n$-sheet partition function to obtain the entanglement entropy for one-interval, which shows a shift of the central charge.

\subsection{Boundary Effective Action on the Sphere Manifold}
\noindent
The bulk Euclidean AdS$_3$ metric can be asymptotically written as $ds_{3a}^2=r^2ds^2_{\mathrm{s}}+dr^2/r^2$,
where $ds^2_{\mathrm{s}}=d\psi^2+\sin^2\psi d\theta^2$, $0\le\psi<\pi$, and $0\le\theta<2\pi$.
The $\psi$ is the Euclidean time defined by $\psi\equiv it$.
The line element $ds^2_{\mathrm{s}}$ is for the unit sphere. The asymptotic behaviors of the gauge fields for the Lorentzian AdS$_3$ metric are: 
\bea
A_{r\rightarrow\infty}=
\begin{pmatrix}
\frac{dr}{2r}& 0
\\
rE^+& -\frac{dr}{2r}
\end{pmatrix},
\qquad
\bar{A}_{r\rightarrow\infty}=
\begin{pmatrix}
-\frac{dr}{2r}& -rE^{-}
\\
0& \frac{dr}{2r}
\end{pmatrix},
\nn\\
\eea
where $E^+\equiv E^{\theta}+E^{t}$ and $E^-\equiv E^{\theta}-E^{t}$
are the boundary zweibein. Then we can find the below boundary condition by replacing $r\rightarrow rE_{\theta}^{\pm}$: 
$\lambda=\sqrt{2rE_{\theta}^+/\partial_{\theta}F}$, $\Psi=-(\partial_{\theta}^2F/\partial_{\theta}F)/(4rE_{\theta}^+)$, $\bar{\lambda}=\sqrt{2rE_{\theta}^-/\partial_{\theta}\bar{F}}$, and $\bar{\Psi}=-(\partial_{\theta}^2\bar{F}/\partial_{\theta}\bar{F})/(4rE_{\theta}^-)$.
For the sphere manifold, we have $E^{\psi}=d\psi$ and $E^{\theta}=\sin\psi d\theta$.
Because we did the Wick rotation, we use the following coordinates: $x^+=-i\psi+\theta$, $x^-=-i\psi-\theta$, $\psi=i(x^++x^-)/2$, and $\theta=(x^+-x^-)/2$.
The $\theta$-component of the boundary zweibein is defined by the $E^{\pm}_{\theta}$. Therefore, we have $E^+_{\theta}=E^-_{\theta}=\sin\psi$.
The boundary gauge-field in the Lorentzian manifold satisfies the conditions: $E_{\theta}^+A^{t}-E_{t}^+A^{\theta}=0$ and $E_{\theta}^-\bar{A}^{t}-E_{t}^-\bar{A}^{\theta}=0$.
Therefore, the AdS$_3$ gravitation action with the spherical asymptotic boundary is \cite{Cotler:2018zff}
\bea
&&S_{\mathrm{GS}}
\nn\\
&=&\frac{k}{2\pi}\int d^3x\ \epsilon^{tr\theta}\mathrm{Tr}\bigg(A_tF_{r\theta}-\frac{1}{2}\big(A_r\partial_tA_{\theta}-A_{\theta}\partial_tA_r\big)\bigg)
\nn\\
&&-\frac{k}{2\pi}\int d^3x\ \epsilon^{tr\theta}\mathrm{Tr}\bigg(\bar{A}_t\bar{F}_{r\theta}-\frac{1}{2}\big(\bar{A}_r\partial_t\bar{A}_{\theta}-\bar{A}_{\theta}\partial_t\bar{A}_r\big)\bigg)
\nn\\
&&+\frac{k}{4\pi}\int dtd\theta\ \mathrm{Tr}\bigg(\frac{E_{t}^+}{E_{\theta}^+}A_{\theta}^2\bigg)
\nn\\
&&-\frac{k}{4\pi}\int dtd\theta\ \mathrm{Tr}\bigg(\frac{E_{t}^-}{E_{\theta}^-}\bar{A}_{\theta}^2\bigg).
\eea
Then we use the conditions $\lambda^2\partial_{\theta}F=2E_{\theta}^+r$ and $\bar{\lambda}^2\partial_{\theta}\bar{F}=2E_{\theta}^-r$ to obtain the boundary effective action on the sphere manifold \cite{Cotler:2018zff}
\bea
S_{\mathrm{GS}}
=\frac{k}{\pi}\int dtd\theta\ \bigg(\frac{(\partial_{\theta}\lambda)(D_-\lambda)}{\lambda^2}
-\frac{(\partial_{\theta}\bar{\lambda})(D_+\bar{\lambda})}{\bar{\lambda}^2}\bigg),
\nn\\
\eea
where 
\bea
D_+\equiv\frac{1}{2} \partial_t+\frac{1}{2}\frac{E_{t}^-}{E_{\theta}^-}\partial_{\theta}, \qquad D_-\equiv\frac{1}{2}\partial_t+\frac{1}{2}\frac{E_{t}^+}{E_{\theta}^+}\partial_{\theta}.
\eea
From the field redefinition: 
${\cal F}\equiv F/E^+_{\theta}$ and $\bar{{\cal F}}\equiv\bar{F}/E_{\theta}^-$, 
the gravitation action on the sphere manifold becomes \cite{Cotler:2018zff}:
\bea
&&S_{\mathrm{GS}}
\nn\\
&=&\frac{k}{4\pi}\int dtd\theta\ \bigg(\frac{(\partial_{\theta}^2{\cal F})(D_-\partial_{\theta}{\cal{F}})}{(\partial_{\theta}{\cal F})^2}
-\frac{(\partial_{\theta}^2\bar{{\cal F}})(D_+\partial_{\theta}\bar{{\cal{F}}})}{(\partial_{\theta}\bar{{\cal F}})^2}\bigg)
\nn\\
&=&\frac{k}{4\pi}\int dtd\theta\ \bigg\lbrack\frac{(\partial_{\theta}^2\phi)(D_-\partial_{\theta}\phi)}{(\partial_{\theta}\phi)^2}
-(\partial_{\theta}\phi)(D_-\phi)\bigg\rbrack
\nn\\
&&
-\frac{k}{4\pi}\int dtd\theta\ \bigg\lbrack\frac{(\partial_{\theta}^2\bar{\phi})(D_+\partial_{\theta}\bar{\phi})}{(\partial_{\theta}\bar{\phi})^2}
-(\partial_{\theta}\bar{\phi})(D_+\bar{\phi})\bigg\rbrack,
\eea
in which we used ${\cal F}\equiv\tan(\phi/2)$ and $\bar{{\cal F}}\equiv\tan(\bar{\phi}/2)$.
Because the boundary theory is scale invariant, we can use the scale transformation to compute the entanglement entropy as in CFT.

\subsection{Entanglement Entropy for One-Interval}
\noindent
Now we want to compute the R\'enyi entropy 
\begin{equation}
\label{eq:REE}
S_n\equiv(\ln Z_n-n\ln Z_1)/(1-n)
\end{equation}
 from the replica trick \cite{Holzhey:1994we} on the $\theta$-direction,
where $Z_n$ is the $n$-sheet partition function, and $Z_1$ is same as the partition function. Because we only consider the computation up to the one-loop correction, we obtain that the partition function is a product of the classical partition-function ($Z_c$) and the one-loop partition-function ($Z_q$) $Z_n=Z_{n, c}\cdot Z_{n, q}$.
When we take the logarithmic on the $n$-sheet partition function, we obtain $\ln Z_n=\ln Z_{n, c}+\ln Z_{n, q}$.
Hence we can treat the classical and one-loop partition-functions separately in the computation of the R\'enyi entropy. Our motivation is to compare the entanglement entropy to the expectation value of the Wilson line. Therefore, we will take the limit $n\rightarrow 1$ in the R\'enyi entropy to obtain the entanglement entropy. Now we compute the $Z_n$ on the sphere manifold, and then the result corresponds to the entanglement entropy for one-interval.
\\

\noindent
We first perform the coordinate transformation to get $ds_s^2=\sech^2(y)(dy^2+d\theta^2)$, in which we used $\sech y=\sin\psi$.
In the $n$-sheet manifold, the range of the $\theta$ is $0<\theta\le 2\pi n$. The periodicity of this theory for $\theta$ is $2\pi n$. 
When we do the computation, we need to regularize the range of the $y$-direction. The range of the $y$-direction is $-\ln(L/\epsilon)< y\le \ln(L/\epsilon)$. The periodicity of this theory for the $y$-direction is $4\ln(L/\epsilon)$ because we assume the Dirichlet boundary condition in the $y$-direction. The $L$ is the length of an interval, and $\epsilon$ is the cut-off on the ending point of the interval. 
\\

\noindent
Finally, we identify the sphere from the torus to determine the complex structure $\tau_{n}$ on the sphere. The coordinates of torus $z\equiv(\theta+iy)/n$
 satisfy the identification: $z\sim z+2\pi $ and $z\sim z+2\pi\tau_{n}$.
The boundary condition of the fields, $\phi$ and $\bar{\phi}$ is given by $\phi(y/n, \theta/n+2\pi )=\phi(y/n, \theta/n)+2\pi$, $\phi\big(y/n+2\pi\cdot\mathrm{Im}(\tau_{n}), \theta/n+2\pi\cdot\mathrm{Re}(\tau_{n})\big)=\phi(y/n, \theta/n)$, $\bar{\phi}(y/n, \theta/n+2\pi)=\bar{\phi}(y/n, \theta/n)+2\pi$, and $\bar{\phi}\big(y/n+2\pi\cdot\mathrm{Im}(\tau_{n}), \theta/n+2\pi\cdot\mathrm{Re}(\tau_{n})\big)=\bar{\phi}(y/n, \theta/n)$.
Therefore, we can quickly find that the complex structure on the sphere is $\tau_{n}=\big(2i/(n\pi)\big)\ln(L/\epsilon)$.
The fields on the sphere can be expanded from the way: $\phi=\theta/n+\epsilon(y, \theta)$ and $\bar{\phi}=-\theta/n+\bar{\epsilon}(y, \theta)$,
where
\bea
\epsilon(y, \theta)&\equiv&\sum_{j, k}\epsilon_{j, k} e^{i\frac{j}{n}\theta-\frac{k}{\tau}y}, \qquad
\epsilon_{j, k}^*\equiv\epsilon_{-j, -k},
\nn\\
\bar{\epsilon}(y, \theta)&\equiv&\sum_{j, k}\bar{\epsilon}_{j, k} e^{i\frac{j}{n}\theta-\frac{k}{\tau}y}, \qquad
\bar{\epsilon}_{j, k}^*\equiv\bar{\epsilon}_{-j, -k},
\eea
and $\tau=n\tau_{n}$.
Because this theory has the SL(2) redundancy, the variables have the constraints: $\epsilon_{j, k}=0$ and $\bar{\epsilon}_{j, k}=0$ when $j=-1, 0, 1$.
To compute the partition function on the sphere, we need to do the Wick rotation $t=-i\psi$,
and then the derivative becomes:
\bea
D_+&=&
\frac{1}{2}\partial_t+\frac{1}{2}\frac{E_{\psi}^-}{E_{\theta}^-}\partial_{\theta}
=-\frac{i}{2}\cosh(y)\partial_y-\frac{1}{2}\cosh(y)\partial_{\theta}, 
\nn\\
D_-&=&
\frac{1}{2}\partial_t+\frac{1}{2}\frac{E_{\psi}^+}{E_{\theta}^+}\partial_{\theta}
=-\frac{i}{2}
\cosh(y)\partial_y+\frac{1}{2}\cosh(y)\partial_{\theta},
\nn\\
\eea
and the gravitation action becomes
\bea
&&S_{\mathrm{GS}}
\nn\\
&=&\frac{k}{4\pi}\int_{\frac{-\pi}{2}\cdot\mathrm{Im}(\tau)}^{\frac{\pi}{2}\cdot\mathrm{Im}(\tau)} dy\int_0^{2\pi n} d\theta\
\nn\\
&&\times \sech(y)
\bigg\lbrack\frac{(\partial_{\theta}^2\phi)(D_-\partial_{\theta}\phi)}{(\partial_{\theta}\phi)^2}
-(\partial_{\theta}\phi)(D_-\phi)\bigg\rbrack
\nn\\
&&
-\frac{k}{4\pi}\int_{\frac{-\pi}{2}\cdot\mathrm{Im}(\tau)}^{\frac{\pi}{2}\cdot\mathrm{Im}(\tau)} dy \int_0^{2\pi n} d\theta\ 
\nn\\
&&\times\sech(y)
 \bigg\lbrack\frac{(\partial_{\theta}^2\bar{\phi})(D_+\partial_{\theta}\bar{\phi})}{(\partial_{\theta}\bar{\phi})^2}
-(\partial_{\theta}\bar{\phi})(D_+\bar{\phi})\bigg\rbrack.
\eea
Substituting the saddle-points into the action, we obtain 
\bea
S_{\mathrm{GS}}=-\frac{c}{6n}\ln\bigg(\frac{L}{\epsilon}\bigg),
\eea
 where $c=6k$ is the central charge of the CFT$_2$ \cite{Brown:1986nw}. Therefore, we obtain $\ln Z_{n, c}=\big(c/(6n)\big)\ln(L/\epsilon)$.
The R\'enyi entropy \eqref{eq:REE} from the saddle-points is given by:
\bea
S_{n, c}=\frac{c}{1-n}\bigg(\frac{1}{6n}-\frac{n}{6}\bigg)\ln\frac{L}{\epsilon}
=\frac{c(1+n)}{6n}\ln\frac{L}{\epsilon},
\eea
in which we used $-n\ln Z_{1, c}=-n(c/6)\cdot\ln(L/\epsilon)$. When we take $n\rightarrow 1$, we obtain $S_{1, c}=(c/3)\ln(L/\epsilon)$.
\\

\noindent
Now we consider the perturbation $\epsilon(y, \theta)$ to obtain the one-loop effect. 
Because the $\phi$-part and $\bar{\phi}$-part are the same, we can only consider the field $\phi$ to obtain the one-loop correction in the R\'enyi entropy. 
The expansion from the $\epsilon$ in the gravitation action for the $\phi$-part is
\bea
&&\frac{k}{4\pi}\int_{\frac{-\pi}{2}\cdot\mathrm{Im}(\tau)}^{\frac{\pi}{2}\cdot\mathrm{Im}(\tau)} dy\int_0^{2\pi n}d\theta\
\nn\\
&&\times \bigg(n^2\big(\partial_{\theta}^2\epsilon(y, \theta)\big)\big(\bar{\partial}\partial_{\theta}\epsilon(y, \theta)\big)-\big(\partial_{\theta}\epsilon(y, \theta)\big)\big(\bar{\partial}\epsilon(y, \theta)\big)\bigg)
\nn\\
&=&-i\frac{k}{8\pi}\sum_{j, k}j(j^2-1)\bigg(k+\frac{j}{n}\tau\bigg)|\epsilon_{j, k}|^2,
\eea
where $\bar{\partial}\equiv(-i\partial_y+\partial_{\theta})/2$.
Therefore, we obtain
\bea
\partial_{\tau}\ln Z_{n, q}=-\sum_{j\neq 0,\pm 1}\sum_{k=-\infty}^{\infty}\frac{\frac{j}{n}}{k+\frac{j}{n}\tau}.
\eea
Now we use the following useful equation $\tilde{\psi}(1-x)-\tilde{\psi}(x)=\pi\cot(\pi x)$,
in which the digamma function is defined by
\bea
\tilde{\psi}(a)\equiv
-\sum_{n=0}^{\infty}\frac{1}{n+a}.
\eea
Therefore, we obtain:
\bea
\sum_{m=-\infty}^{\infty}\frac{1}{m-x}&=&
-\sum_{m=0}^{\infty}\frac{1}{m+x}+\sum_{m=0}^{\infty}\frac{1}{m+1-x}
\nn\\
&=&\tilde{\psi}(x)-\tilde{\psi}(1-x)=-\pi\cdot\cot(\pi x).
\eea
Hence we get
\bea
\partial_{\tau}\ln Z_{n, q}
=-2\pi\sum_{j=2}^{\infty}\bigg(\frac{j}{n}\bigg)\cdot\cot\bigg(\frac{j\pi\tau}{n}\bigg).
\eea
Then we do the re-summation for the above series:
\bea
\partial_{\tau}\ln Z_{n, q}
&=&-2\pi\sum_{j=2}^{\infty}\bigg(\frac{j}{n}\bigg)\cdot\cot\bigg(\frac{j\pi\tau}{n}\bigg)
\nn\\
&=&-2\pi\sum_{j=2}^{\infty}\frac{j}{n}\cdot\bigg\lbrack\cot\bigg(\frac{j\pi\tau}{n}\bigg)+i\bigg\rbrack
+2\pi i\sum_{j=2}^{\infty}\frac{j}{n}.
\nn\\
\eea
By using the regularization
\bea
\sum_{j=1}^{\infty}j\rightarrow -\frac{1}{12},
\eea
we obtain: 
\bea
\partial_{\tau}\ln Z_{n, q}&=&-2\pi\sum_{j=2}^{\infty}\frac{j}{n}\cdot\bigg\lbrack\cot\bigg(\frac{j\pi\tau}{n}\bigg)+i\bigg\rbrack
+2\pi i\sum_{j=2}^{\infty}\frac{j}{n}
\nn\\
&\rightarrow&-2\pi\sum_{j=2}^{\infty}\frac{j}{n}\cdot\bigg\lbrack\cot\bigg(\frac{j\pi\tau}{n}\bigg)+i\bigg\rbrack
-i\frac{13\pi }{6n}.
\eea
After integrating out the $\tau$, we obtain:
\bea
\ln Z_{n, q}=-2\sum_{j=2}^{\infty}\bigg\lbrack\ln\sin\bigg(\frac{\pi j\tau}{n}\bigg)+i\frac{j\pi\tau}{n}\bigg\rbrack-i\frac{13\pi\tau}{6n}+\cdots,
\nn\\
\eea
where $\cdots$ is independent of the $\tau$. The above series is convergent for the $\mathrm{Im}(\tau)>0$.
When we take the limit $L/\epsilon\rightarrow\infty$, we obtain $\ln Z_{n, q}=\big(13/(3n)\big)\ln(L/\epsilon)$.
The R\'enyi entropy for the one-loop correction is: 
\bea
S_{n, q}=\big(\ln Z_{n, q, CS}-n\ln Z_{1, q, CS}\big)/(1-n)=\frac{13(n+1)}{3n}\ln\bigg(\frac{L}{\epsilon}\bigg),
\eea 
where $Z_{n, q, CS}$ is the $n$-sheet partition of the $\phi$ and $\bar{\phi}$.
Therefore, we obtain the R\'enyi entropy $S_{n}=\big((c+26)(n+1)/(6n)\big)\ln(L/\epsilon)$
and the entanglement entropy $S_{EE}=\big((c+26)/3\big)\ln(L/\epsilon)$.
We obtain a shift of the central charge by 26, instead of 13 \cite{Cotler:2018zff}, because $\phi$ and $\bar{\phi}$ each contributes 13, and the sum gives the shift 26.

\section{Wilson Line}
\label{sec:4}
\noindent
The entanglement entropy in the two-dimensional Schwarzian theory gives the conformal deviation from the quantum correction.  Here we want to obtain a bulk description of the entanglement entropy. Since the Wilson lines \cite{deBoer:2013vca}
\bea
W(P, Q)\equiv \mathrm{Tr}\bigg\lbrack{\cal P} \exp\bigg(\int_Q^P\bar{A}\bigg){\cal P}\exp\bigg(\int_{Q}^{P}A\bigg)\bigg\rbrack,
\eea
can provide the entanglement entropy in the CFT$_2$, we begin from this operator to study. The ${\cal P}$ denotes the path-ordering, $P$ and $Q$ are the two-ending points of the Wilson lines at a time slice. Here the trace operation acts on the representation.
\\

\noindent
We extend the Wilson line to the following form \cite{Ammon:2013hba}
\bea
&&W_{\cal R}(C)
\nn\\
&=&\int DU DP D\lambda\ 
\nn\\
&&\times\exp\bigg\lbrack\int_C ds\ \bigg(\mathrm{Tr}(PU^{-1}D_s U)
\nn\\
&&+\lambda(s)\big(\mathrm{Tr}(P^2)-c_2\big)\bigg)\bigg\rbrack,
\eea
where $U$ is an SL(2) element, $P$ is its conjugate momentum, $\sqrt{2c_2}\equiv c(1-n)/6$, and the covariant derivative is defined as that:
$D_sU\equiv dU/ds+A_sU+U\bar{A}_s$ and $A_s\equiv A_{\mu}\cdot(dx^{\mu}/ds)$.
The equations of motion are:
\bea
i\big(k/(2\pi)\big)F_{\mu_1\mu_2}&=&-\int ds\ (dx^{\mu_3}/ds)\epsilon_{\mu_1\mu_2\mu_3}\delta^3\big(x-x(s)\big)UPU^{-1},
\nn\\ 
i\big(k/(2\pi)\big)\bar{F}_{\mu_1\mu_2}&=&\int ds\ (dx^{\mu_3}/ds)\epsilon_{\mu_1\mu_2\mu_3}\delta^3\big(x-x(s)\big)P.
\label{eq:eom}
\eea
A solution of the equations of motion is that \cite{Ammon:2013hba}: $A=g^{-1}ag+g^{-1}dg$, $g=\exp(L_1 z)\exp(\rho L_0)$; 
$\bar{A}=-\bar{g}^{-1}a\bar{g}-\bar{g}^{-1}d\bar{g}$, $\bar{g}=\exp(L_{-1}\bar{z})\exp(-\rho L_0)$,
where the gauge field is given as $a=\sqrt{c_2/2}\cdot (1/k)\cdot(dz/z-d\bar{z}/\bar{z})L_0$.
The SL(2) algebra is defined by that: $\lbrack L_j, L_k\rbrack=(j-k)L_{j+k}$, $j, k=0,\pm 1$; $\mathrm{Tr}(L_0^2)=1/2$, $\mathrm{Tr}(L_{-1}L_1)=-1$, and the traces of other bilinears vanish.
Here we choose $z\equiv r\exp(i\Phi)$ and $\bar{z}\equiv r\exp(-i\Phi)$.
Then the spacetime interval is $ds_3^2=d\rho^2+\exp({2\rho})(dr^2+n^2r^2d\Phi^2)$ \cite{Ammon:2013hba}. 
This solution corresponds to $U(s)=1$, $P(s)=\sqrt{2c_2}L_0$ with the curve $z(s)=0$ and $\rho(s)=s$.
Hence we find that including the Wilson line directly gives the $n$-sheet geometry \cite{Ammon:2013hba}. When this geometry approaches the boundary, it is the $n$-sheet cylinder ($dt^2+n^2d\Phi^2$) up to a scale transformation by using $r\equiv\exp(t)$. 
\\

\noindent
 Let us then comment about the quantum correction of the Wilson line. 
  The backreaction of the Wilson line leads to the $n$-sheet manifold, and only the Chern-Simons term survives when $n\rightarrow 1$, because the right hand sides of \eqref{eq:eom} vanish in that case. 
 Hence this implies that entanglement entropy can be calculated from the Chern-Simons term on the $n$-sheet manifold and the analytical continuation. 
 The expectation value of Wilson line $\langle W_{\cal R}\rangle$ is $Z_n/Z_1^n+\cdots$, where $\cdots$ vanishes when $n\rightarrow 1$, and $Z_n$ is the $n$-sheet partition function of the two-dimensional Schwarzian theory, which is equivalent to the Chern-Simons theory with the same boundary condition. 
 In other words, the entanglement entropy is \cite{Ammon:2013hba}
\bea
S_{EE}=\lim_{n\rightarrow 1}\frac{1}{1-n}\ln \langle W_{\cal R}\rangle,
\eea
where $\langle W_{\cal R}\rangle$ is the expectation value of the Wilson line. Substituting the classical solution of the two-dimensional Schwarzian theory into the Wilson line, it provides the entanglement entropy of CFT$_2$ \cite{deBoer:2013vca}, which implies that the Wilson line can be seen as the geodesic line at the on-shell level. Moreover, the equivalence between the Wilson line and the entanglement entropy is exact. As explained earlier, the reason is that the Chern-Simons partition function $Z_n$ for computing $\langle W_{\cal R}\rangle$ reduces to its boundary version of a Schwarzian theory, the latter of which is essentially the same partition introduced in Sec~\ref{sec:3} (due to the same boundary condition).
Hence the Wilson line can be seen as the appropriate operator to provide one equivalent description of the minimum surface even at the quantum level.   

\section{Outlook}
\label{sec:5}
\noindent
The Wilson line was used in the AdS$_3$ Einstein gravity theory for obtaining the entanglement entropy of CFT$_2$ \cite{Holzhey:1994we} at the classical level \cite{Ammon:2013hba, deBoer:2013vca}. It will be promoted to an operator at a quantum level. We first computed the entanglement entropy in the boundary theory, two-dimensional Schwarzian theory \cite{Cotler:2018zff}. Then we used the Wilson line to obtain the bulk description for the boundary entanglement entropy. This shows that the Wilson line is a suitable operator providing an equivalent description of the minimum surface in the usual correspondence of ``minimum surface=entanglement entropy''. 
\\

\noindent
One should observe that entanglement entropy is related to the expectation value of the Wilson line, not logarithm of the Wilson line. 
Because the logarithm of Wilson is an area operator, the quantum contribution of entanglement entropy will give the non-area term. 
Hence this should justify that area term is not enough for holographic entanglement entropy \cite{Ryu:2006bv}. 
Our study is for the Chern-Simons formulation \cite{Witten:1988hc}. 
Hence our result possibly may not be applied to the metric formulation at a quantum level.
\\

\noindent
The concept of metric is only defined at the classical level but no common sense at the quantum level. 
We know that entanglement entropy in CFT is dual to the AdS minimum surface \cite{Ryu:2006bv}. 
By observing the quantum correction of entanglement entropy, quantum deformation of the minimum surface is given by the fluctuation of a Wilson line or a gauge field, not a metric field. 

\section*{Acknowledgments}
\noindent
We would like to thank Chuan-Tsung Chan, Bartlomiej Czech, Jan de Boer, Kristan Jensen, and Ryo Suzuki for their useful discussion.
\\

\noindent
Xing Huang acknowledges the support of NWU Starting Grant No.0115/338050048 and the Double First-class University Construction Project of Northwest University.
Chen-Te Ma was supported by the Post-Doctoral International Exchange Program and China Postdoctoral Science Foundation, Postdoctoral General Funding: Second Class (Grant No. 2019M652926), and would like to thank Nan-Peng Ma for his encouragement.
Hongfei Shu was supported by the JSPS Research Fellowship 17J07135 for Young Scientists, from Japan Society for the Promotion of Science (JSPS) and the grant “Exact Resultsin Gauge and String Theories” from the Knut and Alice Wallenberg foundation.
 We would like to thank the Shing-Tung Yau Center at the Southeast University, Shanghai University, National Tsing Hua University, Institute for Advanced Study at the Tsinghua University, and Yangzhou University.
 \\
 
 \noindent
Discussions during the workshop, ``East Asia Joint Workshop on Fields and Strings 2019'', was useful to complete this work.


\end{document}